\documentclass[a4paper,fleqn,usenatbib]{mnras}

\usepackage[T1]{fontenc}
\usepackage{ae,aecompl}
\usepackage{ulem,soul}

\usepackage{graphicx}	% Including figure files
\usepackage{amsmath}	% Advanced maths commands
\usepackage{amssymb}	% Extra maths symbols

\usepackage{graphicx,color}
\usepackage[usenames,dvipsnames,svgnames,table]{xcolor}

\title[Stability of the advection-dominated discs]{The local stability of the magnetized advection-dominated discs with the radial viscous force}
\author[Ghoreyshi \& Shadmehri]{S. M. Ghoreyshi$^{1,2}$\thanks{E-mail: smghoreyshi64@gmail.com}\ and M. Shadmehri$^{1}$\\
$^{1}$ Department of Physics, Faculty of Sciences, Golestan University, Gorgan 49138-15739, Iran\\
$^{2}$ Research Institute for Astronomy and Astrophysics of Maragha (RIAAM) Maragha, IRAN, P. O. Box: 55134 - 441}

\begin{document}

%\date{Accepted ???. Received ???; in original form ???}

\pagerange{\pageref{firstpage}--\pageref{lastpage}} \pubyear{2017}
\maketitle

\label{firstpage}

%%%%%%%%%%%%%%%%%%%%%%%%%%%%%%%%%%%%%%%%%%%%%%%%%%%%%%%%%%%%%%%%%%%%%%%%%%%%%%%%%%%%%%%%%%%%%%%%%%%%%%%%%%%%%%
%%%%%%%%%%%%%%%%%%%%%%%%%%%%%%%%%%%%%%%%%%%%%%%%%%%%%%%%%%%%%%%%%%%%%%%%%%%%%%%%%%%%%%%%%%%%%%%%%%%%%%%%%%%%%%

\begin{abstract}
We study local stability of the advection-dominated optically thick (slim) and optically thin discs with purely toroidal magnetic field and the radial viscous force using a linear perturbation analysis. Our dispersion relation indicates that the presence of magnetic fields and radial viscous force cannot give rise to any new mode of the instability. We find, however, that growth rate of the thermal mode in the slim discs and that of the acoustic modes in the slim and optically thin discs are dramatically affected by the radial viscous force. This force tends to strongly decrease the growth rate of the outward-propagating acoustic mode (O-mode) at the short-wavelength limit, but it causes a slim disc to become thermally more unstable. We find that growth rate of the thermal mode increases in the presence of radial viscous force. This enhancement is more significant when the viscosity parameter is large. We also show that growth rate of the O-mode reduces when radial viscous force is considered. The growth rates of the thermal and acoustic modes depend on the magnetic field. Although the instability of O-mode for a stronger magnetic field case has a higher growth rate, the thermal mode of the slim discs can be suppressed when the magnetic field is strong. The inertial-acoustic instability of a magnetized disc may explain the quasi-periodic oscillations (QPOs) from the black holes.
\end{abstract}


%%%%%%%%%%%%%%%%%%%%%%%%%%%%%%%%%%%%%%%%%%%%%%%%%%%%%%%%%%%%%%%%%%%%%%%%%%%%%%%%%%%%%%%%%%%%%%%%%%%%%%%%%%%%%%
%%%%%%%%%%%%%%%%%%%%%%%%%%%%%%%%%%%%%%%%%%%%%%%%%%%%%%%%%%%%%%%%%%%%%%%%%%%%%%%%%%%%%%%%%%%%%%%%%%%%%%%%%%%%

\section{Introduction}

Several galactic and extra galactic radiation sources are believed to be powered by an accretion disc orbiting a very compact object like a black hole (BH) or a neutron star (Shields et al. 2000; Kondratko et al. 2005; Desroches et al. 2009; Yuan et al. 2010). Depending on the accretion rate and the radiative efficiency, different theoretical models have been proposed over recent decades to explain observational features of these complex systems (e.g., see Yuan et al. 2002; Kato et al. 2008; Straub et al. 2011). While the standard accretion disc model (Shakura \& Sunyaev 1973; Pringle 1981) is a convenient model for describing accretion flows around some of these objects, there are significant observational and theoretical arguments that some of the accretion flows are not radiatively efficient (e.g., Narayan \& Yi 1994; Quataert 2001; Chiaberge \& Macchetto 2006; Ho 2009). It was a good incentive for astronomers to construct models for radiatively inefficient accretion flows (RIAFs) which subsequent studies have shown that these models are successful to explain certain observational features of some accreting systems. The advection-dominated accretion flows (ADAFs; Ichimaru 1977, Narayan \& Yi 1994) and slim discs (Abramowicz et al. 1988) are the most widely studied models in this context (for a review, see, e. g., Kato et al. 2008). RIAFs are successful in explaining the spectral transitions of X-ray binaries (Godet et al. 2012), X-ray emission in FR I radio galaxy (Wu et al. 2007), ultraluminous compact X-ray sources (Watarai et al. 2001; Chen \& Wang 2004; Vierdayanti et al. 2006; Soria et al. 2015), strong soft X-ray emission and time variability of narrow-line Seyfert 1 galaxies (Mineshige et al. 2000; Wang \& Netzer 2003; Chen \& Wang 2004).

Instabilities in the accretion discs are believed to be responsible for some observational features such as variabilities in the active galactic nuclei
(AGNs) and the X-ray binaries (Janiuk \& Czerny 2011; Dexter and Quataert 2012; Jiang et al. 2013; Wu et al. 2016), and the periodic X-ray variability
of the Seyfert galaxies (Choi et al. 2002; Czerny et al. 2003; Kawaguchi 2003). The QPOs in BH discs are attributed to the viscous-thermal instability and the inertial-acoustic (pulsational) instability (e.g., Chen \& Taam 1994, 1995; Taam 1999; Dubus et al. 2001; Fan et al. 2008; Grzedzielski et al. 2015). Although various types of the disc instabilities have been studied by many authors (e.g., Abramowicz et al. 1984; Wu et al. 1995a, b, c; Ding
et al. 2000; Khosravi \& Khesali 2014, hereafter KK14), we briefly review the previous works related to the present study.

Thermal instability (TI) has been proposed to account for the dwarf novae burst phenomena (Osaki 1974). TI related to dwarf novae outburst was found by H\={o}shi (1979) and Meyer \& Meyer-Hofmeister (1981). This instability in the discs around BHs or neutron stars is a promising mechanism to explain observed variabilities of X-ray binaries, galactic nuclei, and quasars (Shibazaki \& H\={o}shi 1975; Shakura \& Sanyave 1976; Kakubari et al. 1989; Huang \& Wheeler 1989; Choi et al. 2002; Grzedzielski et al. 2015). Radiation-dominated inner region of a standard accretion disc is subject to TI and viscous instability (Lightman \& Eardley 1974; Shakura \& Sanyave 1976; Piran 1978; Blumentahal et al. 1984; Wu et al. 1995b; Wu \& Li 1996, hereafter WL96; Dubus et al. 2001; Kawanaka \& Kohri 2012; Khesali \& Khosravi 2013, hereafter KK13; Jiang et al.
2013; Tessema 2014; Yu et al. 2015). A gas-pressure-dominated disc, however, is thermally and viscously stable (Wu et al. 1995a). These instabilities can be suppressed by the advective cooling (KK13) and the effects associated with magnetic fields (Begelman \& Pringle 2007; Tessema 2014; S\c{a}dowski 2016).

The pulsational instability can be excited in a viscous disc (Kato 1978; Blumenthal et al. 1984; Wu et al. 1995c). Physical ingredients like polytropic index (Cao \& Zhang 1994), advective cooling (Ding et al. 2000), and magnetic fields (Yang et al. 1995a; Zhou 2013) are able to affect growth rate of this instability. Inclusion of the advective cooling, for instance, bifurcates the acoustic mode into an outward-propagating acoustic mode (hereafter O-mode) and an inward-propagating acoustic mode (hereafter I-mode). Departure of these two modes becomes more significant as the advective cooling dominates over radiative cooling (Wu 1997; Ding et al. 2000). Yang et al. (1995a) also showed that properties of the acoustic modes in the outer region of a magnetized, isothermal thin disc depend on the propagation direction and the ratio of the radial to the azimuthal components of the magnetic field.

Properties of the thermal mode in ADAFs and the slim discs depend on the disc optically depth, the wavelength of perturbation, the accretion rate (Narayan \& Yi 1995; WL96; Kato et al. 1996, 1997; Fujimoto and Arai 1998), heat diffusion and magnetic fields (Yamasaki 1997). Although it has been shown that ADAFs are thermally stable, slim discs are thermally unstable (Kato et al. 1996; WL96). Properties of the acoustic modes in a slim disc are dependent on the viscosity parameter, accretion rate and propagation direction of the perturbations (Wallinder 1995; see also KK13).

Radial viscosity in the ADAFs or slim discs has often been neglected for simplicity, while this quantity has been implemented by a few authors under restricted conditions (e.g., Wu 1997; Ding et al. 2000). Linear stability analysis of a polytropic disc with the radial viscous force indicates that disc stability
depends strongly on the ratio of the radial to the azimuthal viscosities (Wu et al. 1994). The radial viscous force has a stabilizing
influence on the acoustic modes, and this effect is more significant in the outer part of a polytropic disc comparing to its inner region (Wu et al. 1994). Yu et al. (1994) and Yang et al. (1995b) also investigated the effect of radial viscous force on the stability of an isothermal magnetized accretion disc. They showed that the radial component of the viscous force does not give rise to any new unstable mode and damps the radial oscillations. Although the radial viscous force in a non-isothermal disc stabilizes the acoustic modes, it does not modify the thermal and the viscous modes (Chen \& Taam 1993). In this context, role of advection has also been studied by WL96 and KK13. Although they considered the radial viscous force in their basic equations, they did not explore physical consequences of this force. A hot two-temperature ADAF with the radial viscous force and thermal diffusion is viscously and thermally stable (Wu 1997). But such a disc can be thermally unstable subject to the azimuthal perturbations (Ding et al. 2000). Note that the influences of the radial viscous force on the stability of ADAFs were not discussed by Wu (1997) and Ding et al. (2000).

To best of our knowledge, the influence of the radial viscosity on the stability of a magnetized advection-dominated disc has not yet been explored in
detail. Our goal is to examine stability of the advection-dominated discs and investigate properties of the excited modes when the toroidal component of a magnetic field and radial viscous force are considered simultaneously. In section 2, we present basic equations and the associated linearized forms to obtain a general dispersion relation. We then explore stability properties of the excited modes in section 3. We conclude by a summary of our results and possible astrophysical implications in the final section.

%%%%%%%%%%%%%%%%%%%%%%%%%%%%%%%%%%%%%%%%%%%%%%%%%%%%%%%%%%%%%%%%%%%%%%%%%%%%%%%%%%%%%%%%%%%%%%%%%%%%%%%%%%%%%%
%%%%%%%%%%%%%%%%%%%%%%%%%%%%%%%%%%%%%%%%%%%%%%%%%%%%%%%%%%%%%%%%%%%%%%%%%%%%%%%%%%%%%%%%%%%%%%%%%%%%%%%%%%%%%%

\section{General Formulation}

In this section, we present basic equations and their linearized forms to obtain a general dispersion relation which enable us to explore properties of the excited modes.

\subsection{Basic Equations}

We begin by writing the basic equations for a magnetized disc under certain simplifying assumptions. A cylindrical coordinate system $(r,\phi,z)$ is adopted that centered on the central object. We assume that the disc is axisymmetric (i.e., $\partial/\partial \phi=0$) and non-self-gravitating. The toroidal component of the magnetic field is assumed to be dominant and the effects of magnetic diffusivity are neglected. The pseudo-Newtonian potential $\Psi$ is used to describe the gravitational field of the central object (Paczy\'{n}ski \& Wiita 1980), i.e.,
\begin{center}
 $ \Psi=GM/(R-r_{\rm g}),$
\end{center}
where $M$ is the mass of central BH. Here, the distance from the central object is denoted by $R=(r^2+z^2)^{1/2}$ and the Schwarzschild radius is $r_{\rm g}=
2GM/c^2$. The vertical integrated equations, therefore, are written as (e.g., Kato et al. 2008):
\begin{equation}\label{a1}
\frac{\partial\Sigma}{\partial t} + \frac{1}{r} \frac{\partial}{\partial r} (r \Sigma V_r)=0,~~~~~~~~~~~~~~~~~~~~~~~~~~~~~~~~~~~~~~~
\end{equation}
\begin{eqnarray}\label{a2}
\Sigma \frac{\partial V_r}{\partial t} + \Sigma V_r \frac{\partial V_r}{\partial r} - \Sigma (\Omega^2- \Omega_{\rm K}^2)r=~~~~~~~~~~~~~~~~~~~~
 \nonumber\\
-2 \frac{\partial } {\partial r} [H(P+\frac{{B_\phi}^2}{8\pi})]+F_\nu-\frac{H {B_\phi}^2}{2\pi r},~~~
\end{eqnarray}
\begin{equation}\label{a3}
\Sigma r^3 \frac{\partial \Omega}{\partial t} + \Sigma r V_r \frac{\partial}{\partial r} (r^2 \Omega)=\frac{\partial}{\partial r}(\Sigma\nu r^3 \frac{\partial \Omega}{\partial r}),~~~~~~~~~~~~~~~~
\end{equation}
\begin{eqnarray}\label{a4}
\frac{\partial E}{\partial t} - (E+\Pi)\frac{\partial}{\partial t} (\ln\Sigma)+\Pi\frac{\partial}{\partial t}(\ln H)~~~~~~~~~~~~~~~~~~~~~~
\nonumber\\~~~+{Q^-}_{\rm adv}={Q^+}_{\rm vis}+{Q^+}_{\rm vis_{r}}-{Q^-}_{\rm rad}+Q_{\rm t},
\end{eqnarray}
\begin{equation}\label{a5}
\frac{\partial B_\phi}{\partial t} + \frac{\partial}{\partial r} (V_ r B_\phi)=0,~~~~~~~~~~~~~~~~~~~~~~~~~~~~~~~~~~~~~~~
\end{equation}
where $\Sigma$, $V_ r$, $B_\phi$, $\Omega$ and $T$ are the surface density, radial velocity, the toroidal component of the magnetic field, angular velocity and temperature, respectively. The pressure $P$ is defined as $P=\rho {c_{\rm s}}^2 =P_{\rm gas}+P_{\rm rad}$, where the gas pressure is $P_{\rm gas}=\Re\rho T$, and the radiation pressure is $P_{\rm rad}=\frac{1}{3} a T^4$. Here, the volume density, the sound speed, the gas constant, and the radiation constant are denoted by $\rho$, $c_{\rm s}$, $\Re$ and $a$, respectively. Assuming that the disc is in hydrostatic equilibrium in the vertical direction implies that $H=c_{\rm s}/\Omega_{\rm K}$, where $H$ is disc thickness and $\Omega_{\rm K}$ denotes Keplerian angular velocity at the equatorial plane, i.e. $\Omega_{\rm K}=(\frac{1}{r} \frac{\partial \Psi}{\partial r})^{1/2}|_{z=0}$. Contribution of the magnetic pressure to the disc thickness is not included because we are considering the weak field cases. We define the total pressure $P_{\rm tot}$ as sum of the pressure $P$
and the magnetic pressure $P_{\rm mag}={B_\phi}^2/8\pi$. Furthermore, $F_\nu$ is the radial viscous force. The turbulent viscosity $\nu$ associated with the azimuthal direction is defined as $\alpha c_{\rm s} H$ (Shakura \& Sunyaev 1973).

In the energy equation (4), the quantities ${Q^-}_{\rm adv}$, ${Q^+}_{\rm vis}$, ${Q^+}_{\rm vis_{r}}$, ${Q^-}_{\rm rad}$ and $Q_{\rm t}$ are the advection cooling, viscous heating, local viscous dissipation rate associated with stress in the radial direction, radiation cooling and thermal diffusion, respectively. Although Radial viscosity has been considered by WL96, Wu (1997), Ding et al. (2000), KK13 and KK14, the radial viscous heating ${Q^+}_{\rm vis_{r}}$ was neglected by them. The vertical integrated pressure is $\Pi=2HP_{\rm tot}$ and the internal energy $E$ is written as
\begin{center}
 $E
  =[\beta/(\gamma-1)+3(1-\beta-\beta_{\rm m})]\Pi,$
\end{center}
where $\beta=P_{\rm gas}/P_{\rm tot}$, and $\beta_{\rm m}=P_{\rm mag}/P_{\rm tot}$. Since the ratio of the advective cooling to the viscous heating is proportional to the square of the ratio of the disc half-thickness to the radius, $H/r$, the advective energy transport in an advection-dominated disc can dominates over radiative cooling. We can, therefore, safely neglect ${Q^-}_{\rm rad}$ in our model. According to the definition of $Q_{\rm t}$, ${Q^-}_{\rm adv}$, and ${Q^+}_{\rm vis}$, equation (4) is re-written as (see Kato et al. 2008),
\begin{eqnarray}\label{a6}
\frac{\partial E}{\partial t} - (E+\Pi)\frac{\partial}{\partial t} (\ln\Sigma)+\Pi\frac{\partial}{\partial t}(\ln H)~~~~~~~~~~~~~~~~~~~~~
\nonumber\\+V_r[\frac{\partial E}{\partial r} - (E+\Pi)\frac{\partial}{\partial r} (\ln\Sigma)+\Pi\frac{\partial}{\partial r}(\ln H)]
 \nonumber\\ ~~~~~~~~~~~~~~~~~~~~~~=\nu\Sigma(r\frac{\partial \Omega}{\partial r})^2+{Q^+}_{\rm vis_{r}}+\nabla\cdot(K\nabla T),
\end{eqnarray}
where $K$ is the vertical integrated thermal conductivity. The vertical integrated thermal conductivity is written as $K=\alpha f \Omega_{\rm K} H^3 P_{\rm tot}/T$, where $f=3(8-7\beta)f_*$ and $f_*$ is of the order of unity (WL96).

\subsection{Radial Viscosity}

Although the influence of radial viscosity is often neglected in the geometrically thin accretion discs (e.g., Shakura \& Sunyaev 1976; Abramowicz et al. 1984; Cao \&  Zhang 1994; Wu et al. 1995 a, b, c; Chen \& Taam 1995; Tessema 2014), this quantity plays an important role in the inner region of an accretion
disc and in the advection-dominated discs (Wu 1997; Ding et al. 2000; KK13; KK14). Radial component of the stress tensor cannot be ignored when the radial viscosity is significant. The radial viscous force is (Papaloizou \& Stanley 1986):
\begin{equation}\label{a7}
F_\nu= \frac{\partial}{\partial r} [\frac{4}{3}\frac{\nu_r\Sigma}{r}\frac{\partial}{\partial r}(r V_r)]-\frac{2V_r}{r}\frac{\partial}{\partial r}(\nu_r\Sigma),~~~~~~~~~~~~~~~~
\end{equation}
where $\nu_r$ is the kinetic viscosity acting in the radial direction and we assume that $\nu_r =\nu$. One can show that ratio of the radial viscous force to the pressure gradient is proportional to the square of $H/r$. The radial viscous force, therefore, plays a significant role in ADAFs and slim discs with $H/r\leq1$. Next physical quantity that we consider in our analysis is ${Q^+}_{vis_r}$ which can be written as (Chen \& Taam 1993),
\begin{equation}\label{a8}
{Q^+}_{vis_r}=2\nu_r\Sigma\{(\frac{\partial V_r}{\partial r})^2+(\frac{V_r}{r})^2-\frac{1}{3}[\frac{1}{r}\frac{\partial}{\partial r}(r V_r)]^2\}.~~~~
\end{equation}

\subsection{Dispersion Relation}

We are now in a position to investigate local stability of the magnetized ADAFs or slim discs with the radial viscous force. The local approximation means that the wavelength of perturbation $\lambda$ is much smaller than $r$ which can be written as $kr\gg1$. Here, $k$ is the wavenumber of perturbations. So, the local analysis for the advection-dominated discs can be expressed as $\lambda\ll H \leq r$. The validity of vertically integrated equations also requires that $k V_r <\Omega_{\rm K}$. Since the radial velocity $V_r$ is proportional to $\alpha {c_{\rm s}}^2 /r \Omega_{\rm K}$ (Kato et al. 2008), we then obtain
\begin{equation}\label{a9}
\frac{r}{H}
  >\frac{\lambda}{H}>2\pi\alpha\frac{H}{r}.
\end{equation}
These inequalities are satisfied in the discs with $H/r \leq 1$, if the viscosity parameter $\alpha$ is adopted sufficiently small. We, thus, assume that $\alpha=0.01$ and 0.001. These values, for instance, are consistent with the  estimation of the viscosity parameter based on the AGNs observations (Siemiginowska \& Czerny 1989).

The equilibrium state of a disc is constructed using basic equations if we neglect time derivative terms. To obtain the perturbed equations, we
split each variable into unperturbed and perturbed components which are indicated with a subscript $0$ and $1$, respectively. The radial perturbations are in the form $\exp(i(\omega t-kr))$, where $\omega$ is the growth rate of perturbations. The linearized equations, therefore, are written as
\begin{equation}\label{a10}
\tilde{\sigma}\frac{\Sigma_1}{\Sigma_0} -i \frac{\epsilon}{\tilde{H}} \frac{V_{r1}}{r\Omega_{\rm K}}=0,~~~~~~~~~~~~~~~~~~~~~~~~~~~~~~~~~~~~~~~~~
\end{equation}
\begin{eqnarray}\label{a11}
-\{\alpha^2\tilde{H}^4+(\tilde{\Omega}^2-1)+\frac{1}{1+\beta}[2i\alpha^2\epsilon\tilde{H}^3(3\beta-1)~~~~~~~~~
\nonumber\\+2\tilde{V_{\rm a}}^2(1-\beta)-\frac{i\epsilon\tilde{V_{\rm a}}^2(1-\beta)}{2\tilde{H}}-\frac{i\epsilon\beta\tilde{V_{\rm a}}^2}{\tilde{H}(1-\beta_{\rm m})}~~~~~~~~
\nonumber\\~~-i\epsilon\tilde{H}(1-\beta)+\frac{2i\epsilon\beta\tilde{H}}{1-\beta_{\rm m}}]\}\frac{\Sigma_1}{\Sigma_0}+ (\tilde{\sigma} +\frac{1}{3}\alpha\tilde{H}^2+\frac{4}{3}\alpha\epsilon^2
 \nonumber\\~~~+\frac{4}{3}i\alpha\epsilon\tilde{H})\frac{V_{r1}}{r\Omega_{\rm K}}-2\tilde{\Omega}
\frac{\Omega_1}{\Omega_{\rm K}}-\frac{4-3\beta-4\beta_m}{1+\beta}[-2\tilde{V_{\rm a}}^2~~~
 \nonumber\\ +4i\alpha^2\epsilon\tilde{H}^3+\frac{i\epsilon\tilde{H}(2-\beta_m)}{1-\beta_{\rm m}}-\frac{i\epsilon\beta_{\rm m}\tilde{V_{\rm a}}^2}{2\tilde{H}(1-\beta_m)}]\frac{T_1}{T_0}~~
 \nonumber\\~~~+\{3\tilde{V_{\rm a}}^2-\frac{i\epsilon\tilde{V_{\rm a}}^2}{2\tilde{H}}+\frac{1}{1+\beta}[-2i\epsilon\beta_m\tilde{H}+4\beta_m\tilde{V_{\rm a}}^2
\nonumber\\-8i\alpha^2\epsilon\beta_m\tilde{H}^3-\frac{i\epsilon\beta_m\tilde{V_{\rm a}}^2}{\tilde{H}}+\frac{2i\epsilon\beta\beta_{\rm m}\tilde{H}}{1-\beta_{\rm m}}~
\nonumber\\-\frac{i\epsilon\beta\beta_m\tilde{V_{\rm a}}^2}{\tilde{H}(1-\beta_{\rm m})})]\}\frac{{B_\phi}_1}{{B_\phi}_0}=0,
\end{eqnarray}
\begin{eqnarray}\label{a12}
\{\frac{\alpha\tilde{\chi}^2\tilde{H}^2}{2\tilde{\Omega}}+\frac{3\beta-1}{1+\beta}[-2\alpha g \tilde{H}^2 +i\alpha\epsilon g \tilde{H}]\}\frac{\Sigma_1}{\Sigma_0} ~~~~~~~~~~~~
 \nonumber\\+\frac{\tilde{\chi}^2}{2\tilde{\Omega}}\frac{V_{r1}}{r\Omega_k}+(\tilde{\sigma}+2\alpha\tilde{H}^2+3i\alpha\epsilon\tilde{H}+\alpha\epsilon^2)
\frac{\Omega_1}{\Omega_{\rm K}}
 \nonumber\\~~~~~~~+\frac{4-3\beta-4\beta_{\rm m}}{1+\beta}[2i\alpha\epsilon g\tilde{H}-4\alpha g \tilde{H}^2]\frac{T_1}{T_0}
 \nonumber\\~~~~~~~~~~~~~~~~~~~~~~+\frac{2\beta_{\rm m}}{1+\beta}[2i\alpha\epsilon g \tilde{H}-4\alpha g \tilde{H}^2]\frac{{B_\phi}_1}{{B_\phi}_0}=0,
\end{eqnarray}
\begin{eqnarray}\label{a13}
-\frac{1}{1+\beta}[2\tilde{\sigma}(4-3\beta-3\beta_{\rm m})+(\alpha g^2+4\alpha^3\tilde{H}^4)(3\beta~~~~~
 \nonumber\\-1)]\frac{\Sigma_1}{\Sigma_0}+(\frac{\alpha g^2 q}{m \tilde{H}}-4i\alpha^2\epsilon\tilde{H})\frac{V_{r1}}{r\Omega_{\rm K}}+\frac{2i\alpha\epsilon g}{\tilde{H}}\frac{\Omega_1}{\Omega_{\rm K}}
 \nonumber\\+\{[\frac{2\beta(4-3\beta-4\beta_{\rm m})}{(\gamma-1)}-\beta(4-3\gamma)\frac{(7-7\beta-8\beta_{\rm m})}{(\gamma-1)}
  \nonumber\\+(4-3\beta-4\beta_{\rm m})(7-6\beta-3\beta_{\rm m})]\frac{\tilde{\sigma}}{1+\beta}-2\alpha (g^2
 \nonumber\\ +4\alpha^2\tilde{H}^4)\frac{(4-3\beta-4\beta_{\rm m})}{1+\beta}+\frac{1}{2}\alpha\epsilon^2f\}\frac{T_1}{T_0}+\frac{2}{1+\beta}[\tilde{\sigma}\beta_m
 \nonumber\\(4-3\beta-3\beta_{\rm m})-2\alpha\beta_{\rm m} g^2-8\alpha^3\beta_{\rm m}\tilde{H}^4]\frac{{B_\phi}_1}{{B_\phi}_0}=0,
\end{eqnarray}
\begin{equation}\label{a14}
-i \frac{\epsilon}{\tilde{H}} \frac{V_{r1}}{r\Omega_{\rm K}}+(\tilde{\sigma}-\alpha\tilde{H}^2)\frac{{B_\phi}_1}{{B_\phi}_0}=0,~~~~~~~~~~~~~~~~~~~~~~~~~
\end{equation}
where $\tilde{\sigma}=\sigma/\Omega_k$, $\sigma=i(\omega-{\rm K} {V_r}_0)$, $\epsilon=k H$, $\tilde{H}=H/r$, $\tilde{\Omega}=\Omega/\Omega_{\rm K}$,
$g={\tilde{\chi}}^2/2\tilde{\Omega}-2\tilde{\Omega}$, $\tilde{V_{\rm a}}=V_{\rm a}/r\Omega_{\rm K}$, and $\tilde{\chi}=\chi/\Omega_{\rm K}$. Furthermore, $\chi$ is the epicyclic frequency, i.e., $\chi=\{2\Omega_0 [2\Omega_0+r(\partial \Omega_0/\partial r)]\}^{1/2}$. Here, $q$ is the ratio of the advective energy to the dissipated viscous energy, $V_a={B_\phi}_0/\sqrt{4\pi\rho_0}$ is the Alfv\'{e}n velocity and $m$ is the Mach number which is defined as $m={V_r}_0/c_{\rm s}$. Using definitions of $V_{r0}$ and $c_s$, the Mach number becomes $m=\alpha \tilde{H}$ which is dependent on the viscosity parameter. The quantities with a subscript $1$ are the amplitude of the perturbations.

The requirement for the existence of a nontrivial solution for the algebraic equations (10)-(14) (i.e., the determinant of the coefficients vanishes) leads to the following dispersion relation:
\begin{equation}\label{a15}
a_0 \tilde{\sigma}^5+a_1 \tilde{\sigma}^4+a_2 \tilde{\sigma}^3+a_3 \tilde{\sigma}^2+a_4 \tilde{\sigma}+a_5=0.~~~~~~~~~~~~
\end{equation}
This equation can be solved numerically to find unstable modes. The real part of a root corresponds to growth rate of the instability, and
$\nu=\mid {\rm Im}(\tilde{\sigma})\mid\Omega_{\rm K}$ is the corresponding instability frequency. We, however, note that the imaginary part corresponds to the propagating properties. The imaginary part of the O-mode is negative, whereas it is positive for the I-mode. A perturbation grows exponentially when $\tilde{\sigma}$ is real and positive. If $\tilde{\sigma}$ is real and negative, the perturbation damps. Note that $\tilde{\sigma} = 0$ is also a trivial solution. In the non-magnetized case, our dispersion relation (15) reduces to a fourth order polynomial which is in agreement with Chen \& Taam (1993) and WL96. But our dispersion relation (15) is fifth order with respect to growth rate and the increase of order is due to consideration of toroidal magnetic field. When the magnetic fields, the radial viscous force, and the thermal diffusion are neglected, our dispersion relation reduces to the obtained dispersion relation by Fujimoto \& Arai (1998).

%%%%%%%%%%%%%%%%%%%%%%%%%%%%%%%%%%%%%%%%%%%%%%%%%%%%%%%%%%%%%%%%%%%%%%%%%%%%%%%%%%%%%%%%%%%%%%%%%%%%%%%%%%%%%%
%%%%%%%%%%%%%%%%%%%%%%%%%%%%%%%%%%%%%%%%%%%%%%%%%%%%%%%%%%%%%%%%%%%%%%%%%%%%%%%%%%%%%%%%%%%%%%%%%%%%%%%%%%%%%%

\section{Analysis of the Dispersion Relation}

Using dispersion relation (15), we can now explore stability of ADAFs and slim discs for different sets of the input parameters. The dispersion relation is analyzed for $\beta_{\rm m} =0$ (non-magnetized case), 0.1 and 0.15. In all magnetized cases, the non-dimensional Alfv\'{e}n
velocity is $\tilde{V_{\rm a}=0.1}$. The adopted values for $\tilde{H}$ are 0.6 and 0.8. Using equation (9) with $\alpha=0.001$ and $\tilde{H}=0.6$, therefore, the range of $\lambda/H$ becomes between 0.004 and 1.667. If we set $\alpha=0.01$ and $\tilde{H}=0.6$, then $\lambda/H$ varies from 0.04 to 1.667. For $\tilde{H}=0.8$, however, the ratio $\lambda/H$ is between 0.005 and 1.25 for $\alpha=0.001$, and it varies from 0.05 to 1.25 for $\alpha=0.01$. So, one can conclude that ${\rm Re}(\tilde{\sigma})$ is proportional to $\alpha$.

Since the Mach number $m$ depends on the parameters $\alpha$ and $\tilde{H}$, we obtain
\[
\mathrm{if} \ \tilde{H} = 0.6 \ \Rightarrow
m = \begin{cases}
0.006,\ ~~~\ \alpha = 0.01 \\
0.0006, \ ~~\ \alpha = 0.001,
            \end{cases}
\]
and
\[
\mathrm{if} \ \tilde{H} = 0.8 \ \Rightarrow
m = \begin{cases}
0.008, \ ~~~\ \alpha = 0.01 \\
0.0008, \ ~~\ \alpha = 0.001.
            \end{cases}
\]
For the above values of $m$, contribution of the radial velocity in our model can be safely neglected. Our analysis, therefore, is applicable
only to the outer region of a disc where its rotation profile is Keplerian and we have $\tilde{\Omega}=\tilde{\chi}=1$. Thus, we obtain $g=-3/2$. We, however, note that the local approximation is violated in the inner region with $\tilde{\chi}=0$ and $m=1$. Although our focus is to explore the influence of the magnetic fields and the radial viscous force on the disc stability, we also investigate possible effects of the other input parameters.

%%%%%%%%%%%%%%%%%%%%%%%%%%%%%%%%%%%%%%%%%%%%%%%%%%%%%%%%%%%%%%%%%%%%%%%%%%%%%%%%%%%%%%%%%%%%%%%%%%%%%%%%%%%%%%
%%%%%%%%%%%%%%%%%%%%%%%%%%%%%%%%%%%%%%%%%%%%%%%%%%%%%%%%%%%%%%%%%%%%%%%%%%%%%%%%%%%%%%%%%%%%%%%%%%%%%%%%%%%%%%

\subsection{Stability Analysis of the ADAFs}

\begin{figure}
\begin{center}
\includegraphics{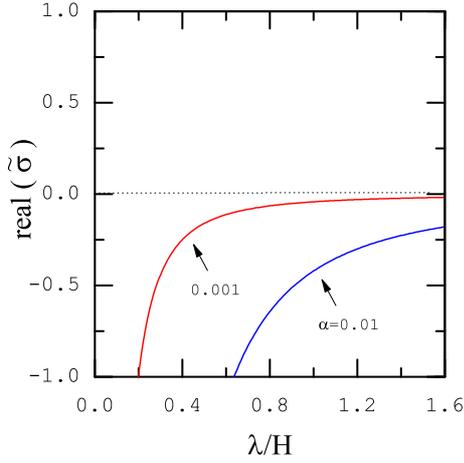}
\end{center}
\vspace{6cm} \caption{The viscous mode of an ADAF for $\alpha=$0.001 and 0.01. This mode becomes more stable with increasing the viscosity parameter $\alpha$.}
\end{figure}

\begin{figure}
\begin{center}
\includegraphics{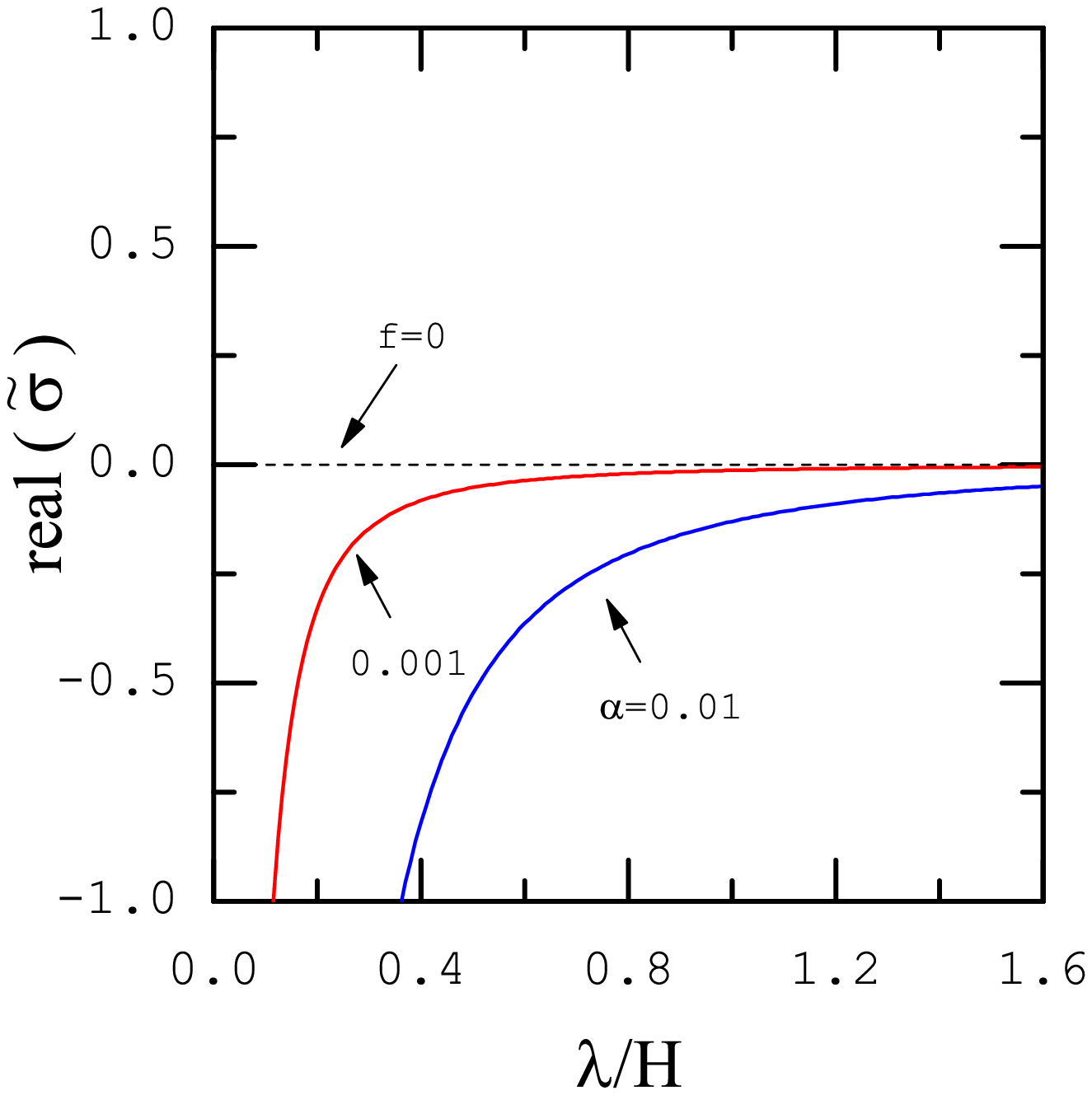} \includegraphics{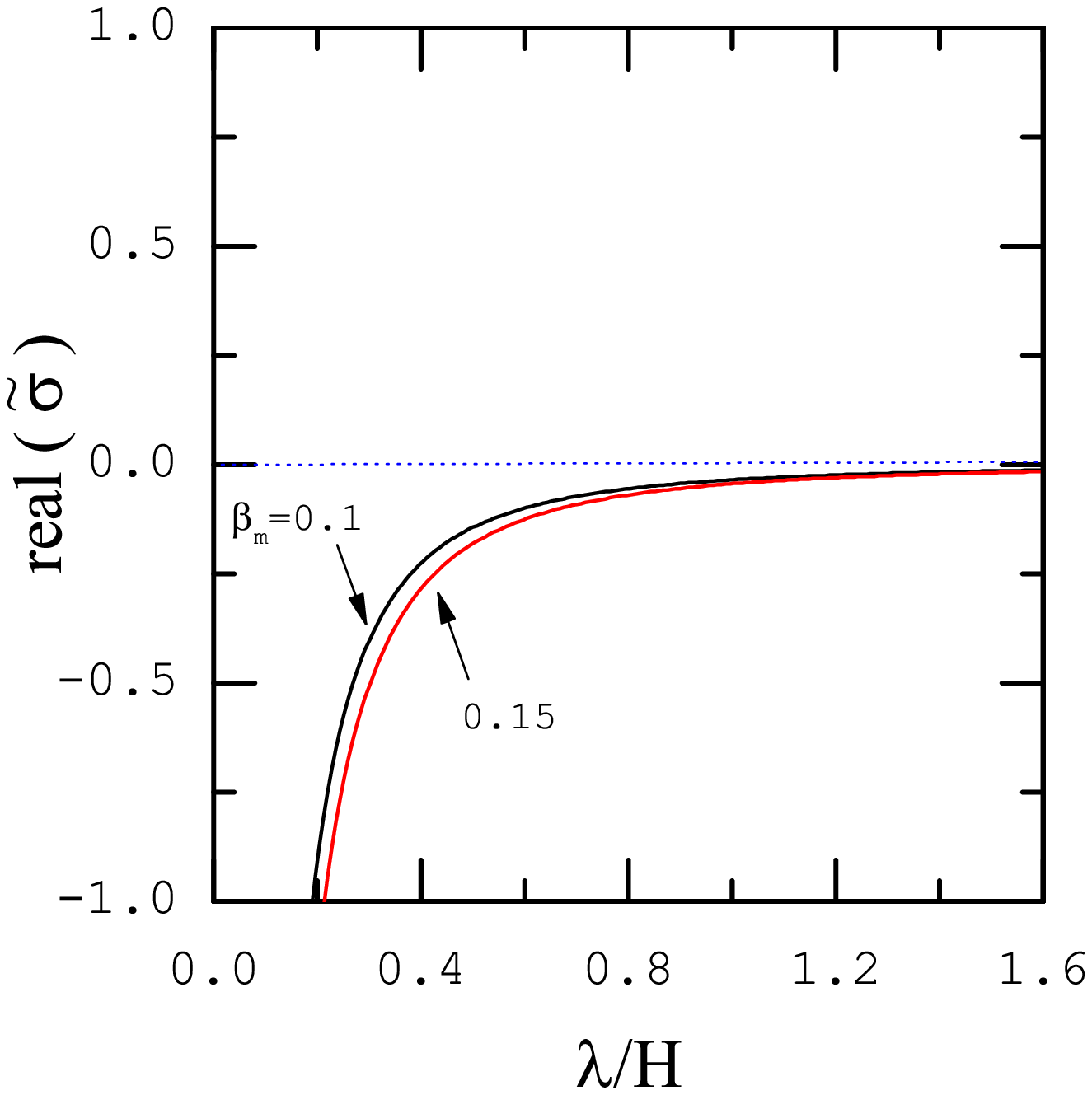}
\end{center}
\vspace{12.6cm} \caption{Top panel displays the influence of the viscosity parameter $\alpha$ on the growth rate of the thermal mode in a non-magnetized ADAF with and without the thermal diffusion. In the bottom panel, the effect of $\beta_m$ on the thermal stability is shown. Here, we have $\alpha=0.001$.}
\end{figure}

In ADAFs, the adopted values of $\beta$ and $\gamma$ are 1.0 and 5/3, respectively. Our general dispersion relation in both non-magnetized and magnetized cases displays four modes of instability; one thermal mode, one viscous mode, and two acoustic modes. We, thereby, confirm that the magnetic fields and the radial viscous force do not give rise to any new mode of the instability. Our analysis shows that the thermal and viscous modes in the ADAFs are always stable which is consistent with previous findings (WL96; KK13; KK14). We also find that the radial viscous force and the disc thickness cannot change the growth rate of the thermal and viscous modes. Among four modes, the viscous mode of a magnetized disc is the same as the viscous mode of a non-magnetized disc. This means that the viscous mode is independent of $\beta_m$, whereas KK14 showed that the viscous mode of an optically thin, advection-dominated disc is modified in the presence of magnetic fields.

\begin{figure}
\begin{center}
\includegraphics{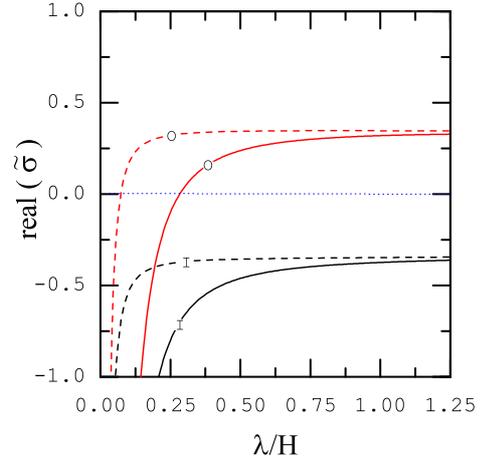} \includegraphics{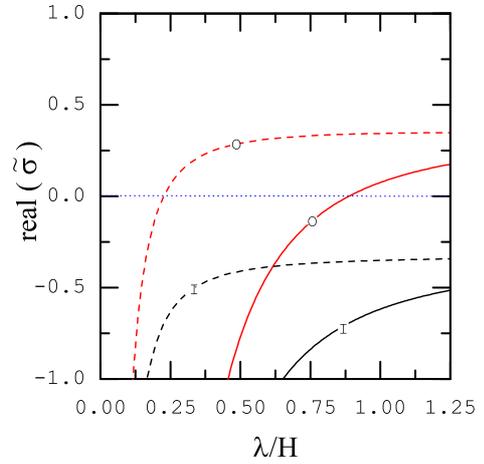}
\end{center}
\vspace{12.6cm} \caption{The growth rate of the acoustic modes of a non-magnetized ADAF with $\tilde{H}=0.8$ and $q=0.99$. The solid and dashed curves are corresponding to the cases with and without the radial viscous force, respectively. Top panel displays the growth rate of ADAFs for $\alpha=0.001$, whereas in the bottom panel, the growth rate of ADAFs is shown for $\alpha=0.01$. The growth of unstable acoustic modes is prevented by the radial viscous force.}
\end{figure}

\begin{figure}
\begin{center}
\includegraphics{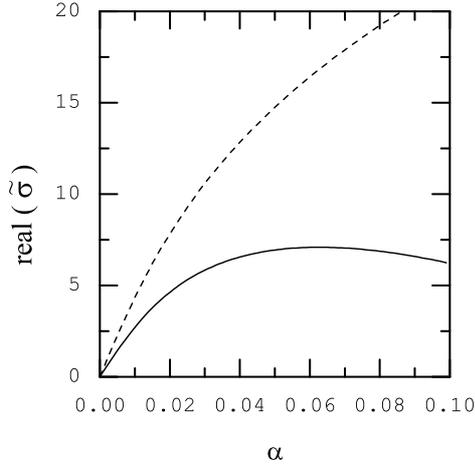}
\end{center}
\vspace{6cm} \caption{The growth rate of O-mode of a non-magnetized ADAF versus $\alpha$, if $\lambda=0.4 H$ and $q=0.99$. The solid and
dashed curves are cases with and without the radial viscous force, respectively.}
\end{figure}

In Figure 1, normalized growth rate of the viscous mode, ${\rm Re} (\tilde{\sigma})$, versus normalized wavelength of the perturbations, $\lambda/H$, is shown for different values of the input parameters. The red and blue solid curves are labeled by the appropriate viscosity parameters. Our analysis shows that the viscous mode is dependent only on the viscosity parameter and becomes more stable as this parameter increases. Figure 2 (top) shows that this parameter has a similar effect on the growth rate of the thermal mode. In addition to the viscosity parameter, thermal diffusion and magnetic fields are able to affect growth rate of the thermal mode. In top panel of Figure 2, the black dashed line corresponds to the case without the thermal diffusion. We find that the thermal mode is excited only in the presence of thermal diffusion. Our results show that the thermal diffusion has an influence neither on the viscous mode nor on the acoustic modes. The magnetic fields are able to modify growth rate of the thermal mode. Figure 2 (bottom) exhibits that the disc tends to be more stable when the ratio $\beta_{\rm m}$ increases. For a given growth rate, one can find that the curve of thermal mode for a higher ratio $\beta_{\rm m}$ is shifted toward longer wavelengths. In other words, the thermal mode has the same growth rate for different values of $\beta_{\rm m}$, if the wavelength of perturbation is different.

\begin{figure}
\begin{center}
\includegraphics{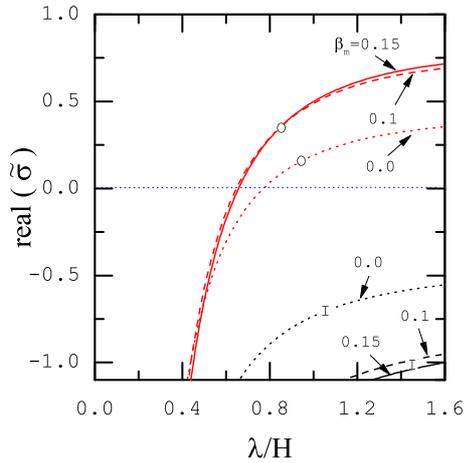}
\end{center}
\vspace{6cm} \caption{The growth rate of the acoustic modes of a magnetized ADAF, if $\tilde{H}=0.6$, $q=0.99$ and $\alpha=0.01$. The red and the black curves are the O-mode and the I-mode, respectively. Each curve is labeled by the appropriate $\beta_m$. Growth rate of the O-mode increases in the presence of magnetic fields.}
\end{figure}

We find that the radial viscous force affects only the acoustic modes of an ADAF. In Figure 3, we display the acoustic modes
in a non-magnetized disc with $\tilde{H}=0.8$ and $\alpha=0.001$ (top) and 0.01 (bottom). The red and black curves are the O-mode and I-mode, respectively.
The cases with and without the radial viscous force are illustrated by solid and dashed curves, respectively. In the explored cases, the I-mode is found to be always stable. The previous works also showed that the I-mode in an ADAF is stable (WL96; KK13). In the short-wavelength limit, the O-mode can also be stable. Although the O-mode in a case with $\alpha=0.001$ is unstable at almost all wavelengths, this mode for $\alpha=0.01$ is unstable in long-wavelength limit. Note that WL96 and KK13 have found stable O-mode in ADAFs.

Figure 3 shows that the presence of radial viscous force causes the curves of the growth rate of acoustic modes to shift toward a longer wavelength when a given growth rate is studied. This force leads to a lower growth rate and to a more stable disc. A similar trend has also been found by Wu et al. (1994) for the O-mode in a polytropic disc. The normalized growth rate of the acoustic modes, ${\rm Re} (\tilde{\sigma})/\alpha$, in our work is larger than finding of Wu (1997) for the same input parameters which means that our O-mode is more unstable. The difference is mainly due to the fact that Wu (1997) considered a two-temperature disc with different heating and cooling functions. Our results show that the influence of the radial viscous force on the growth rate of the acoustic modes depends on the perturbation wavelength. In the short-wavelength perturbation limit, the influence is stronger comparing to the long-wavelength perturbation case. More precisely, the effect of radial viscous force can be negligible at long wavelengths. The change in the viscosity parameter $\alpha$ can also affect the influence of this force. For higher values of $\alpha$, the effect of the radial viscous force is more stronger and the O-mode becomes more stable at longer wavelengths. This trend is illustrated in Figure 4 which shows that growth rate has a much steeper slope in the case without the radial viscous force.

\begin{figure}
\begin{center}
\includegraphics{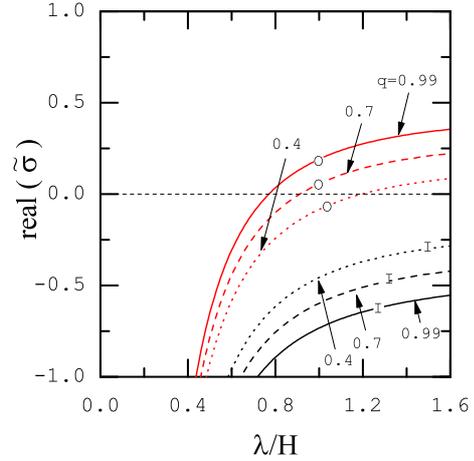}
\end{center}
\vspace{6cm} \caption{The influence of advection on the stability of a non-magnetize ADAF for $\tilde{H}=0.6$ and $\alpha=0.01$. The advection affects only the acoustic modes.}
\end{figure}

The influence of the ratio $\beta_{\rm m}$ on an ADAF with $\tilde{H}=0.6$ is illustrated in Figure 5. The acoustic modes in a magnetized disc have higher growth rates. Thus, the O-mode in the ADAFs with a strong magnetic field is more unstable. Although this mode can be unstable in the long-wavelength limit, KK14 showed that the O-mode is stable for different values of $\beta_m$. Comparison of Figures 3, 5 and 6 shows that the acoustic modes also depend on the values of $\tilde{H}$ and $q$. The growth rate of the O-mode decreases as a disc becomes thicker. We find that the parameter $q$ can only affect the acoustic modes. In a case with a significant advection, the departure of the acoustic modes becomes higher and the growth rate of O-mode is enhanced.

%%%%%%%%%%%%%%%%%%%%%%%%%%%%%%%%%%%%%%%%%%%%%%%%%%%%%%%%%%%%%%%%%%%%%%%%%%%%%%%%%%%%%%%%%%%%%%%%%%%%%%%%%%%%%%
%%%%%%%%%%%%%%%%%%%%%%%%%%%%%%%%%%%%%%%%%%%%%%%%%%%%%%%%%%%%%%%%%%%%%%%%%%%%%%%%%%%%%%%%%%%%%%%%%%%%%%%%%%%%%%

\subsection{Stability analysis of the slim discs}

\begin{figure}
\begin{center}
\includegraphics{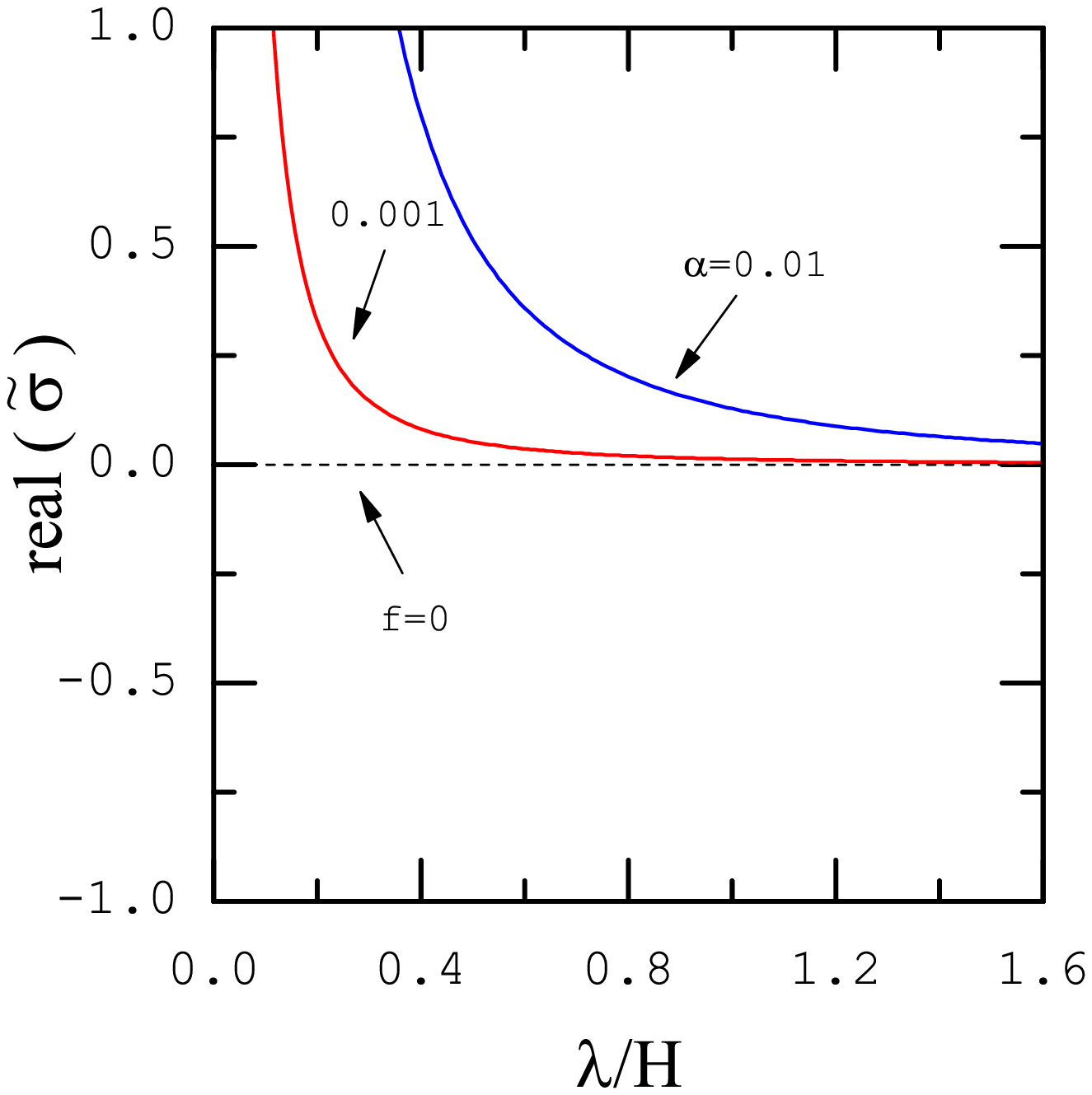} \includegraphics{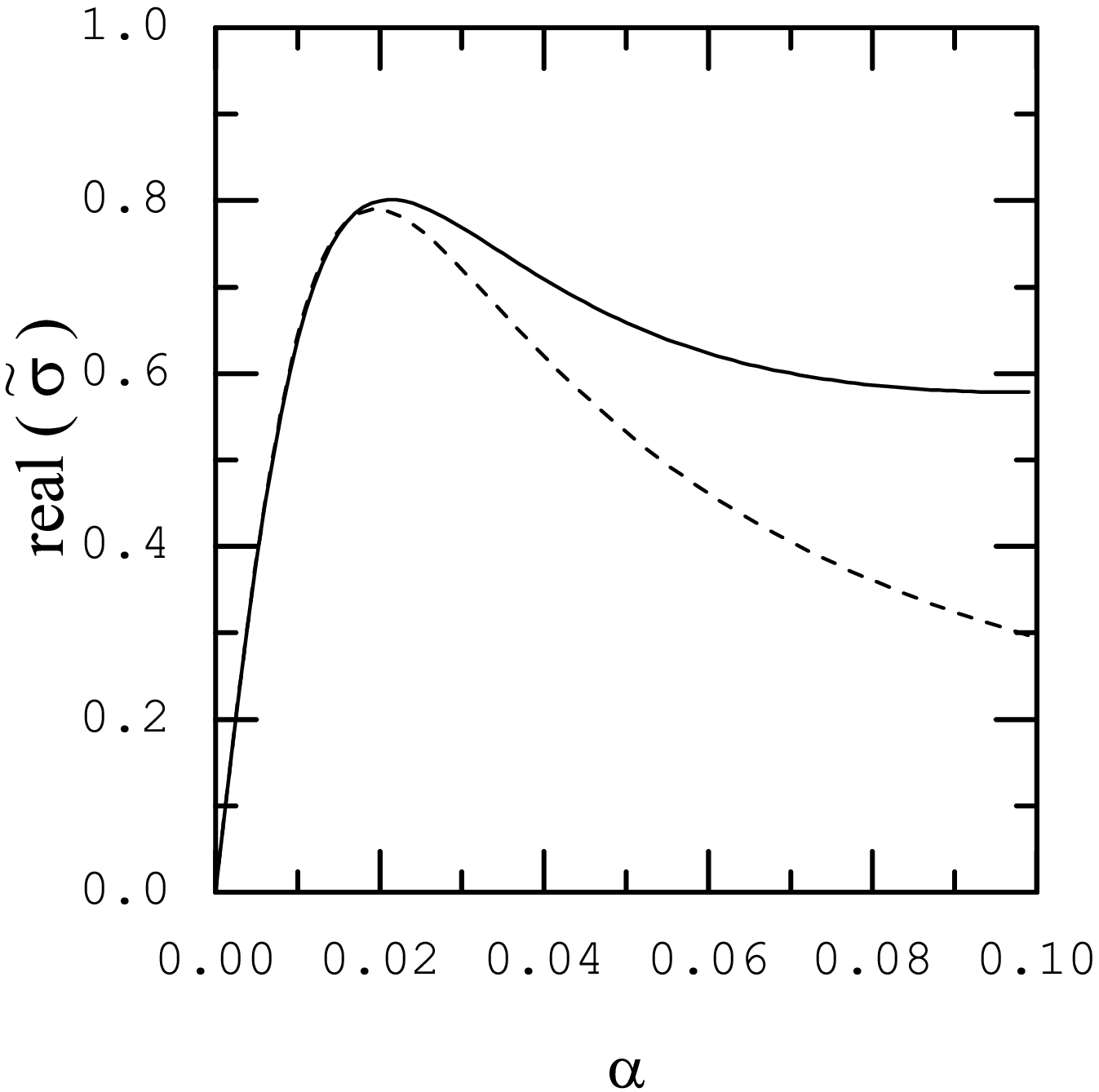}
\end{center}
\vspace{12.6cm}\caption{The thermal mode of a non-magnetized slim disc with $\tilde{H}=0.6$. Top panel shows growth rate of this mode versus wavelength of the perturbation for different values of the viscosity parameter. Bottom panel exhibits growth rate of the thermal mode at a given wavelength $\lambda=0.4 H$ as a function of the viscosity parameter. Here, the solid and dashed curves are corresponding to the cases with and without the radial viscous force, respectively. For higher values of $\alpha$, the thermal mode is more unstable when the radial viscous force is considered.}
\end{figure}

\begin{figure}
\begin{center}
\includegraphics{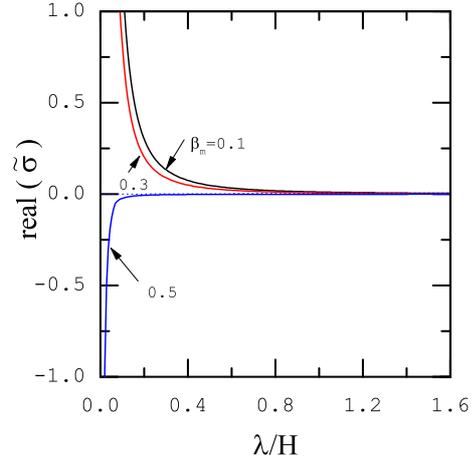}
\end{center}
\vspace{6cm} \caption{Growth rate of the thermal mode of a magnetized slim disc versus perturbation wavelength for $\tilde{H}=0.6$ and $\alpha=0.001$. Each curve is marked by the corresponding value of $\beta_{\rm m}$. The thermal mode becomes stable once the ratio $\beta_{\rm m}$ is larger than 0.5.}
\end{figure}

\begin{figure}
\begin{center}
\includegraphics{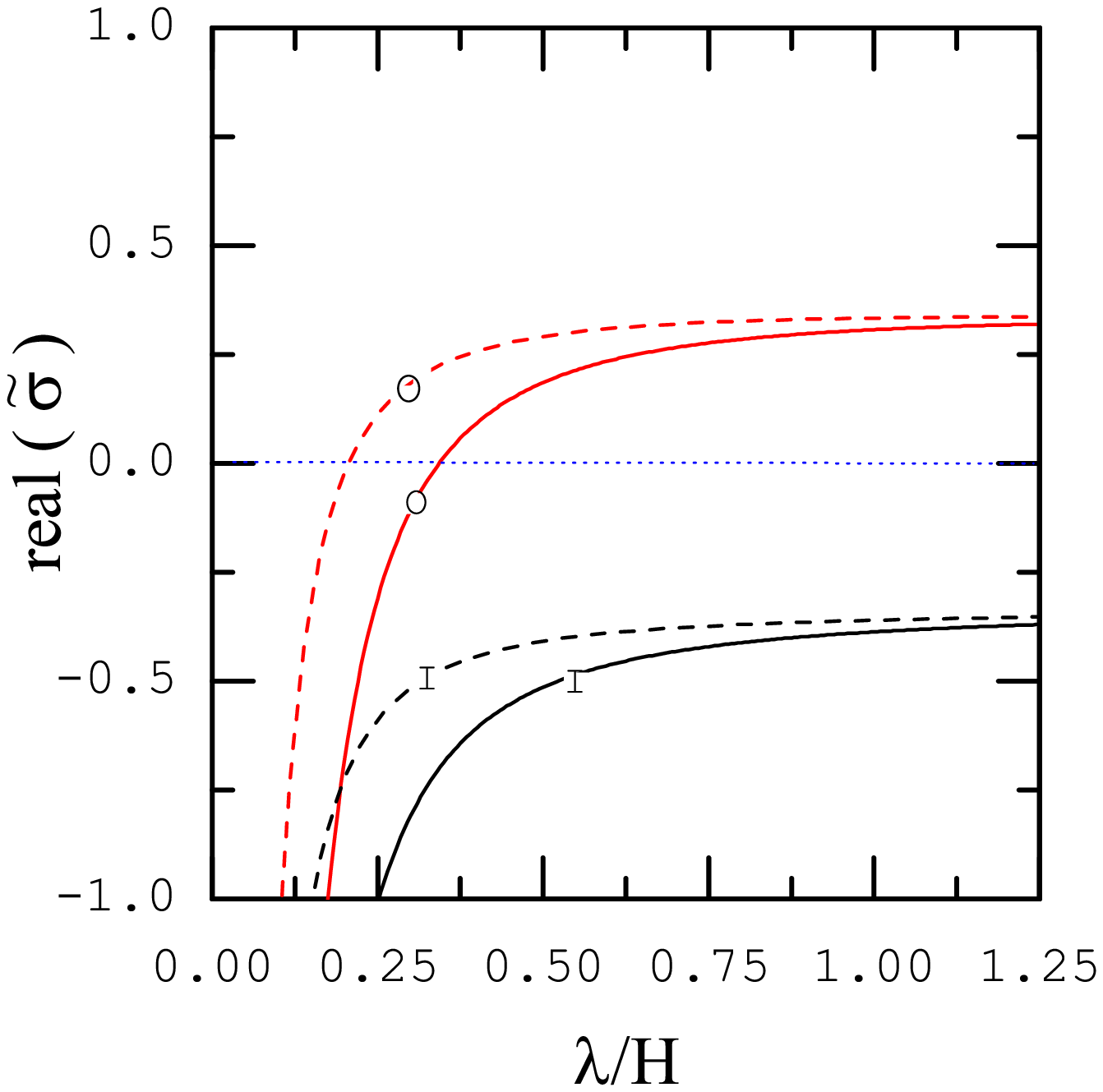} \includegraphics{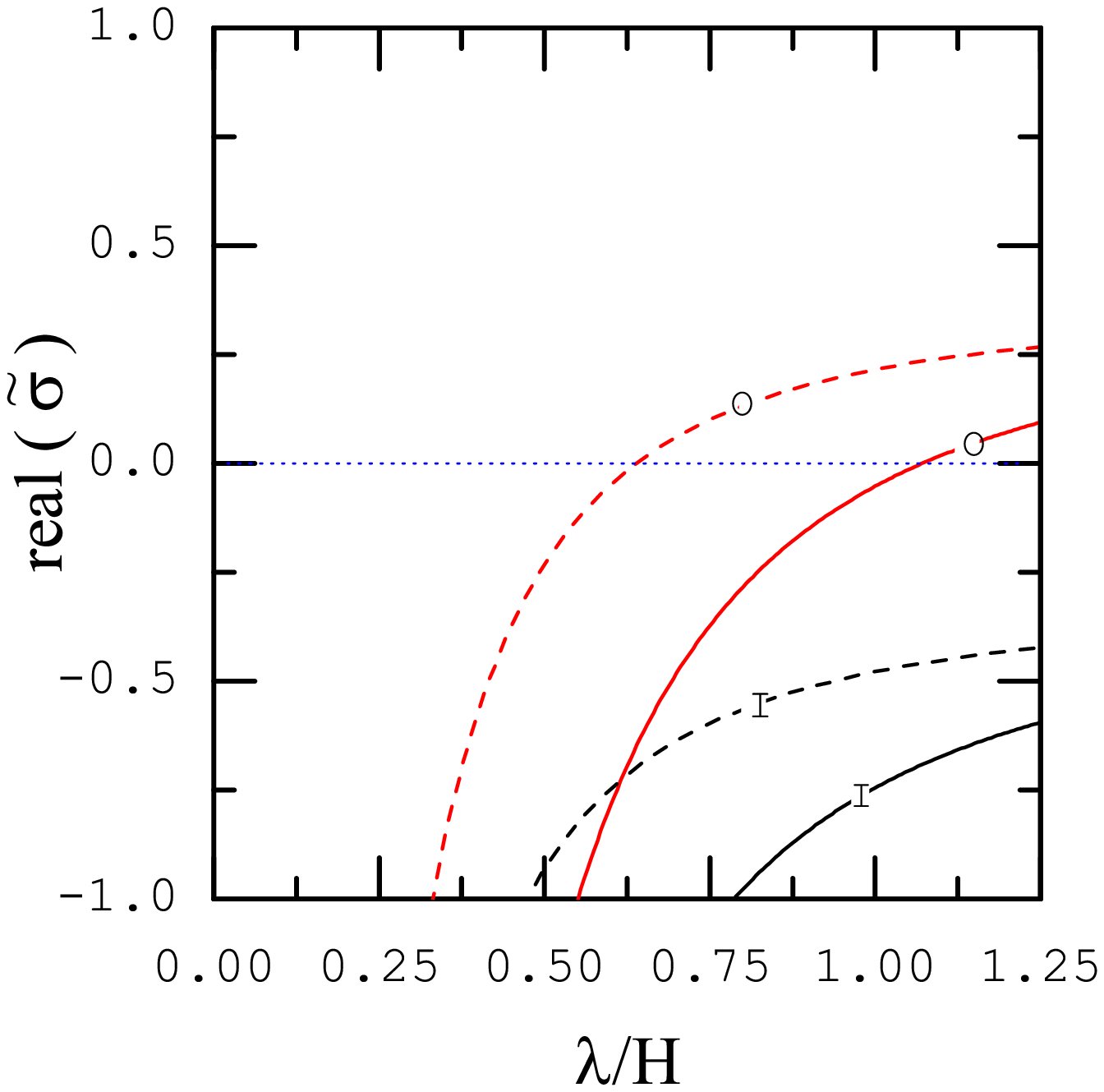}
\end{center}
\vspace{12.6cm} \caption{Growth rate of the pulsational instability as a function of perturbation wavelength in a non-magnetized slim disc with $\tilde{H}=0.8$ and $q=0.99$. The solid and dashed curves are corresponding to the cases with and without the radial viscous force, respectively. Top panel displays the
growth rate for $\alpha=0.001$, whereas in the bottom panel, the growth rate is shown for $\alpha=0.01$. The presence of radial viscous force causes the O-mode for $\alpha=0.01$ to be stabilized at almost all wavelengths.}
\end{figure}

\begin{figure}
\begin{center}
\includegraphics{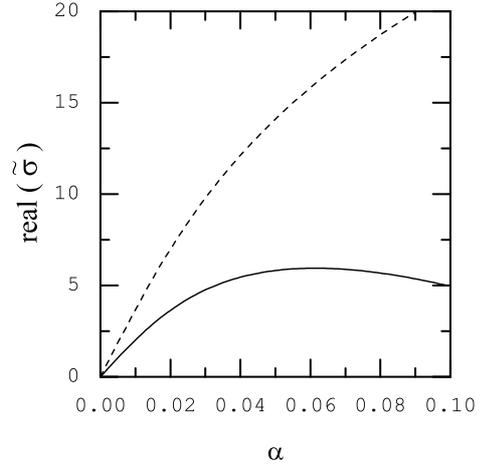}
\end{center}
\vspace{6cm} \caption{Growth rate of the O-mode in a non-magnetized slim disc versus $\alpha$ for $\lambda=0.4 H$ and $q=0.99$. The solid and
dashed curves are cases with and without the radial viscous force, respectively. For higher values of $\alpha$, the influence of the radial viscous force becomes stronger.}
\end{figure}

In the radiation-dominated slim discs, we have $\beta=0$ and $\gamma=4/3$. In this case, the viscous mode is always stable and is dependent only on the viscosity parameter. We find that the stability properties of this mode is independent of the disc optical depth and its growth rate is similar to that of an optically thin disc. The optical depth of disc, however, strongly affects the stability properties of the thermal mode. Figure 7 displays growth rate of the thermal mode as a function of the perturbation wavelength (top), and of viscosity parameter at a given wavelength $\lambda=0.4 H$ (bottom). In top panel of Figure 7, the red and the blue curves are the thermal mode for $\alpha=$0.001 and 0.01, respectively. Growth rate of this mode decreases rapidly with increasing the perturbation wavelength. We also find that the thermal diffusion only affects the thermal mode. A similar trend has also been found in ADAFs. Figure 7 (top) shows that in the absence of the thermal diffusion (the black dashed curve), TI is suppressed which is in agreement with the previous studies
(e.g., WL96). Furthermore, growth rate of the thermal mode is enhanced with increasing the viscosity parameter. Although the radial viscous force in a slim disc with $\alpha=0.001$ and 0.01 is unable to significantly modify the thermal mode, Figure 7 (bottom) shows that this mode is more unstable in the presence of radial viscous force and for higher values of $\alpha$. Our analysis, therefore, shows that thermal mode of a non-magnetized slim disc is always unstable and the presence of radial viscous force implies a higher growth rate.

Role of the magnetic fields on the thermal mode is explored in Figure 8 which shows growth rate of this mode as a function of the perturbation wavelength. Each curve is labeled by the corresponding value of $\beta_{\rm m}$. The curve of thermal mode shifts toward shorter wavelengths with increasing the ratio $\beta_{\rm m}$. Contrary to the finding of KK14, we find that the thermal mode is still unstable for $\beta_{\rm m} =0.3$. Our analysis, however, suggests that the disc will be thermally stable if the ratio $\beta_{\rm m}$ exceeds a value around 0.5. We also find that disc thickness and advection parameter are not able to modify growth rate of TI.

\begin{figure}
\begin{center}
\includegraphics{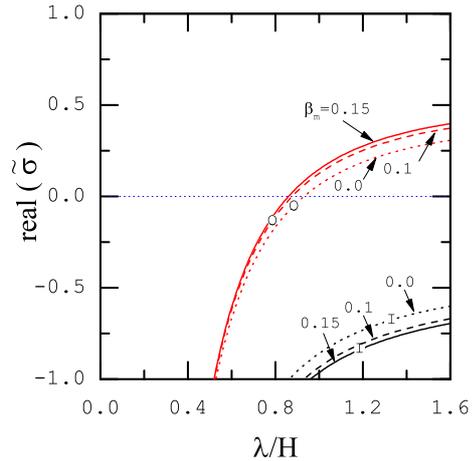}
\end{center}
\vspace{6cm} \caption{Growth rate of the pulsational instability versus perturbation wavelength in a magnetized slim disc for
$\tilde{H}=0.6$, $q=0.99$ and $\alpha=0.01$. The red and the black
curves are the O-mode and the I-mode, respectively. Each curve is labeled by
the appropriate $\beta_m$.}
\end{figure}

We find that the thermal and acoustic modes are affected by the radial viscous force and the magnetic fields. For our adopted viscosity parameter, the effect of radial viscous force on the acoustic modes, in comparison to the other modes, is more significant. In Figure 9, the acoustic modes of a non-magnetized disc are shown for $\tilde{H}=0.8$, $q=0.99$, $\alpha=0.001$ (top), and $\alpha=0.01$ (bottom). The solid and dashed curves are corresponding to the cases with and without the radial viscous force, respectively. In agreement with the finding of KK13 for a slim disc, we also confirm that the I-mode is stable at all wavelengths. Although KK13 found that the O-mode in a slim disc is stable, our analysis shows that this mode can be unstable. The O-mode of a slim disc becomes unstable at a longer wavelength comparing to an ADAF. The radial viscous force also causes the growth rates of the acoustic modes in both optically thick and thin discs to decrease. So, the radial viscous force has a stabilizing role on the acoustic modes, especially at the short wavelengths. Figure 10 shows the viscosity parameter dependence of the growth rate of the O-mode for a given wavelength. This dependence is more significant for large values of $\alpha$ and it is is similar to what we have already obtained for the ADAFs.

Effects of the magnetic fields and the advection parameter on growth rate of the acoustic modes are illustrated in Figures 11 and 12, respectively.
In the presence of magnetic fields, the acoustic modes grow faster. This effect becomes more significant as the ratio $\beta_{\rm m}$ increases.
If $\beta_{\rm m}$ increases, the departure of the two acoustic modes becomes more significant. We also find that the changes of growth rate of the acoustic modes for a slim disc due to the magnetic fields is less than that for an optically thin disc. One can also see that a thicker disc has a lower growth rate. Note that the pulsational instability in thicker discs with $\alpha=0.01$ and in discs with lower advection parameter is nearly damped.

\begin{figure}
\begin{center}
\includegraphics{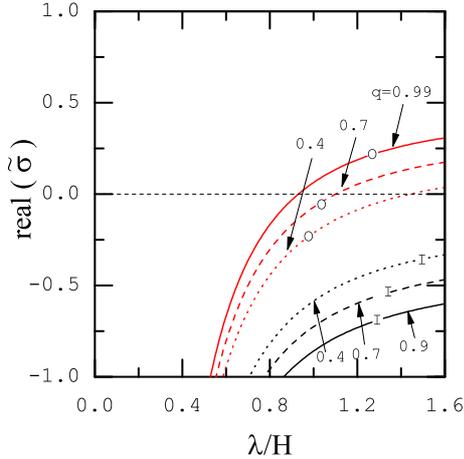}
\end{center}
\vspace{6cm} \caption{The influence of the advection on the stability of a non-magnetize slim disc for $\tilde{H}=0.6$ and $\alpha=0.01$. Only the acoustic modes in a slim disc are affected by the advection.}
\end{figure}

%%%%%%%%%%%%%%%%%%%%%%%%%%%%%%%%%%%%%%%%%%%%%%%%%%%%%%%%%%%%%%%%%%%%%%%%%%%%%%%%%%%%%%%%%%%%%%%%%%%%%%%%%%%%%%
%%%%%%%%%%%%%%%%%%%%%%%%%%%%%%%%%%%%%%%%%%%%%%%%%%%%%%%%%%%%%%%%%%%%%%%%%%%%%%%%%%%%%%%%%%%%%%%%%%%%%%%%%%%%%%

\section{Summary and Discussion}

In this paper we investigated the local stability of the advection-dominated discs with a toroidal magnetic field and the radial viscous force. Using
a linear perturbation method, we found that the presence of magnetic fields and this force cannot introduce any new mode of the instability.
Our results show that the viscous mode is always stable and is independent of the optical depth, the advection parameter, the magnetic field, the disc thickness, and the radial viscous force. However, the growth rates of this mode and of the other modes are strongly dependent on the viscosity parameter. When this parameter increases, the growth rates of acoustic modes are reduced, whereas Wu et al. (1994) showed that the growth rates of these modes rapidly increase with the viscosity parameter. In addition to the dependence of growth rate of the thermal mode on the viscosity parameter, the growth rate of the thermal mode in the slim discs decreases rapidly with increasing perturbation wavelength. We also showed that in the absence of thermal diffusion, TI can be disappeared.

Our numerical calculations suggest that the stability properties and the growth rates of all modes except the viscous mode can be affected by the optical depth. When the disc is optically thin, only the O-mode can be unstable. If the disc is assumed to be optically thick, we found that both the O-mode and the thermal mode are unstable. The growth rates of acoustic modes are not same for cases with different optical depth. Note that the I-mode is always stable. The influence of the radial viscous force on the thermal and the acoustic modes is also dependent on the optical depth of disc. This force can affect the thermal mode of a slim disc and the acoustic modes of ADFs and slim discs. We found that the radial viscous force tends to stabilize the pulsational instability, while it causes the growth rate of TI of a slim disc to increase by a factor of two for $\alpha=0.1$. These findings can be explained using energy dissipation of the wave propagation. We also showed that the influence of the radial viscous force becomes less significant when the wavelength of the perturbation increases.

The magnetic fields can also play an important role in the stability properties of the thermal and the acoustic modes. We found that the slim discs are still thermally unstable when the ratio of magnetic pressure to total pressure is equal to 0.3, which is not in agreement with previous results. But if this ratio has a value around 0.5 or even greater, then the magnetic field can conspicuously suppress TI. So, TI of an optically thick disc can be suppressed due to the presence of a strong magnetic field. Growth rate of the O-mode, however, is increased because of the magnetic fields. We found that the maximum growth rate of an optically thin disc is enhanced by about of factor 2 compared to the non-magnetized mode, while the maximum growth rate of an optically thick disc is about $25\%$ higher.

As mentioned earlier, the viscosity parameter, the thermal diffusion and even the radial viscous force play significant roles in the stability properties of the thermal mode. These quantities can significantly affect heating of the disc and eventually lead to an increase in the temperature. Hence, the disc loses its thermal equilibrium and becomes thermally unstable. The thermally unstable discs undergo a limit cycle which can explain the periodic outbursts of the cataclysmic variables and certain observational features of X-ray transients. The thermal instability can cause the time variations (not only X-ray fluctuation but also periodic light variation) of an accretion disc around a BH. The microquasar GRS 1915+105, for example, displays variability which seems to be well interpreted as an outcome of the thermal instability of a radiation-dominated disc. Of course, some of X-ray binaries with luminosities within the range $0.06L_{\rm E} \leq L \leq 0.5L_{\rm E}$ display very little variability and probably they are stable (e.g., Done et al. 2007). In the light of our findings one can expect that stability of theses systems may be due to the presence of strong magnetic fields.

We also suggest that the inertial-acoustic instability of an advection-dominated optically thin or thick disc may explain the higher frequency QPOs in BH candidates. Theoretical arguments show that pulsation period in the X-ray binaries and AGNs is about $\simeq \alpha^{-1}{\Omega_{\rm K}}^{-1}=4.5\times10^{-4} \alpha^{-1} (M/M_{\bigodot})~s$ (Fan et al. 2008). For $\alpha=0.1$ and $M=10 M_{\bigodot}$, therefore, the obtained period is about $0.045~s$ which is too short to be compared with the observed frequencies ($3-10~Hz$) of the QPOs. Although our results show that radial viscous force may increase the period based on the O-mode, this increment does not give a value consistent with the observations. In the magnetized case, on the other hand, we found that the period increases by a factor of about three for $\alpha=0.1$. The corresponding frequency, thereby, becomes approximately $7~Hz$ which is within the observed interval. We, therefore, speculate that the pulsational instability of a magnetized advection-dominated disk may explain the QPOs observed period in the X-ray binaries and AGNs.

This work has been supported financially by the Research Institute for Astronomy
\& Astrophysics of Maragha (RIAAM) under research project No. 1/5237-61. We are grateful to referee for a detailed report that greatly helped us to improve the paper.

%%%%%%%%%%%%%%%%%%%%%%%%%%%%%%%%%%%%%%%%%%%%%%%%%%%%%%%%%%%%%%%%%%%%%%%%%%%%%%%%%%%%%%%%%%%%%%%%%%%%%%%%%%%%%%
%%%%%%%%%%%%%%%%%%%%%%%%%%%%%%%%%%%%%%%%%%%%%%%%%%%%%%%%%%%%%%%%%%%%%%%%%%%%%%%%%%%%%%%%%%%%%%%%%%%%%%%%%%%%%%

\end{document}